\documentclass[twoside]{ae100prg}
\usepackage{graphicx}
\usepackage[breaklinks]{hyperref}
\usepackage{booktabs}

\begin{document}

\title{Lagrangian analysis of `trivial' symmetries in models of gravity}
\author{Debraj Roy$^1$}
\address{$^1$ S. N. Bose National Centre for Basic Sciences, Block-JD, Sector III, Salt Lake, Kolkata-700098, India.}
\email{debraj@bose.res.in}

\begin{abstract}
We study the differences between Poincar\'{e} and canonical hamiltonian symmetries in models of gravity through the corresponding Noether identities and show that they are equivalent modulo trivial gauge symmetries.
\end{abstract}

\section{Introduction}

Poincar\'{e} symmetry is a fundamental symmetry of nature and a gauge theory of the Poincar\'{e} group can be used to model theories of gravity. This Poincar\'{e} gauge theory (PGT) was developed by Utiyama \cite{Utiyama}, Kibble \cite{Kibble}, Sciama \cite{Sciama} and later on by various authors \cite{Hehl}. PGT is built on a global manifold with local orthonormal frames glued to each spacetime point by frame fields or triads (in 3D). The triads $b^i_{\ \mu}$ are used to translate between the global (index: Greek) and local (index: Latin) frames. To construct a gauge theory, connections $\omega^i_{\ \mu}$ are introduced replacing partial derivatives by corresponding covariant derivatives. The corresponding field strengths give rise to the gravitational fields of curvature $R^i_{\ \mu\nu}$ and torsion $T^i_{\ \mu\nu}$
\begin{eqnarray}
R^i_{\ \mu\nu} &=& \partial_\mu \omega^i_{\ \nu} - \partial_\mu \omega^i_{\ \nu} + \epsilon^i_{\ jk}\,\omega^j_{\ \mu}\omega^k_{\ \nu} \\
T^i_{\ \mu\nu} &=& \nabla_\mu b^i_{\ \nu} - \nabla_\nu b^i_{\ \mu}.
\label{PGT R T}
\end{eqnarray}
These fields can now be used to write actions describing gravity in Riemann-Cartan spacetime. Imposition of a condition on torsion through equations of motion (in vacuum) may lead one to a spacetime with only curvature and no torsion -- the usual Einstein GR on Riemannian manifold.

As gauge theories of the Poincar\'{e} group, Poincar\'{e} symmetries are already inbuilt. A Dirac canonical analysis of symmetries on the other hand also yield a set of gauge symmetries for the same models. By a gauge symmetry here we mean any continuous symmetry of the basic fields that leave the action invariant. The total number of independent gauge symmetries are however limited by the number of independent, primary first class constraints \cite{HTZ}. So it transpires that there is a discrepancy with established results in the apparent off-shell in-equivalence between the Poincar\'{e} and canonical hamiltonian symmetries. Here we study and resolve this from a lagrangian point of view.

\section{Noether identities and trivial symmetries}

For specifics of discussion, we take up the Mielke-Baekler model \cite{MB} describing a cosmologically topological model of gravity with torsion. The action for the model is
\begin{eqnarray}
S = \int \textrm{d$^3$x}\,\epsilon^{\mu\nu\rho}\!\left[ab^i_{\ \mu}R_{i\nu\rho} - \frac{\Lambda}{3} \epsilon_{ijk}b^i_{\ \mu}b^j_{\ \nu}b^k_{\ \rho} + \alpha_3 \left(\omega^i_{\ \mu}\partial_\nu\omega_{i\rho}\right.\right. \nonumber\\
\left.\left.  + \frac{1}{3} \epsilon_{ijk}\,\omega^i_{\ \mu}\omega^j_{\ \nu}\omega^k_{\ \rho} \right) + \frac{\alpha_4}{2}b^i_{\ \mu}T_{i\nu\rho} \right]
\label{action}
\end{eqnarray}
where the terms are the Einstein-Cartan term, cosmological term, Chern-Simons term (in connection) and the torsion term respectively. The Euler derivatives corresponding to the independent canonical fields are:
\begin{eqnarray}
\frac{\delta S}{\delta b^i_{\ \mu}} &=& \epsilon^{\mu\nu\rho} \left[ a\,R_{i\nu\rho} + \alpha_4\, T_{i\nu\rho} - \Lambda\, \epsilon_{ijk}b^j_{\ \nu}b^k_{\ \rho} \right] \nonumber\\
\frac{\delta S}{\delta \omega^i_{\ \mu}} &=& \epsilon^{\mu\nu\rho} \left[ \alpha_3\, R_{i\nu\rho} + a\, T_{i\nu\rho} + \alpha_4\, \epsilon_{ijk}b^j_{\ \nu}b^k_{\ \rho} \right]
\label{EulerDs}
\end{eqnarray}

The model independent Poincar\'{e} symmetries (subscript `$P$') are \cite{BlagoMB}
\begin{eqnarray}
\delta_P b^i_{\ \mu} &=& -\epsilon^i_{\ jk}b^j_{\ \mu}\theta^k - \partial_\mu \xi^\rho \,b^i_{\ \rho} - \xi^\rho\,\partial_\rho b^i_{\ \mu} \nonumber\\
\delta_P \omega^i_{\ \mu} &=& -\partial_\mu \theta^i - \epsilon^i_{\ jk}\omega^j_{\ \mu}\theta^k - \partial_\mu\xi^\rho\,\omega^i_{\ \rho} - \xi^\rho\,\partial_\rho\omega^i_{\ \mu}.
\label{PGT symms}
\end{eqnarray}
while the canonical symmetries generated by the first-class gauge generator constructed through an off-shell algorithm \cite{HTZ, BRR1, BRR2} are \cite{RBDR1}
\begin{eqnarray}
\delta_H b^i_{\ \mu} &=& \nabla_\mu\varepsilon^i - p \,\epsilon^i_{\ jk} \,b^j_{\ \mu} \varepsilon^k + \epsilon^i_{\ jk}\,b^j_{\ \mu} \tau^k \nonumber\\
\delta_H \omega^i_{\ \mu} &=& \nabla_\mu \tau^i - q \,\epsilon^i_{\ jk} \,b^j_{\ \mu} \varepsilon^k.
\label{canonsymm}
\end{eqnarray}

An inspection of the two symmetries (\ref{PGT symms}) and (\ref{canonsymm}) reveal that the canonical symmetries are structurally dependent on the form of the action while the Poincar\'{e} symmetries are independent of particular action. Also, to compare the two symmetries, we have to first find a suitable mapping between the different sets of gauge parameters. To find this, we take recourse to the Noether identities corresponding to the symmetries \cite{RBDR4}.

A Noether identity corresponds to a each continuous gauge symmetry of an action, marked by an independent gauge parameter. Infact, the identity is a direct consequence of the invariance of the action. To see this, let us consider a generic gauge symmetry expressed as terms proportional to the gauge parameter ($\varepsilon^\mu$) and its derivative
\begin{equation}
\delta q_i = R_{i\mu}\varepsilon^\mu + \tilde{R}_{i\mu}^\nu \,(\partial_\nu \varepsilon^\mu).
\label{genGsymm}
\end{equation}
The invariance of the action, step by step, leads to
\begin{eqnarray}
\delta S &=& \int \frac{\delta \mathcal{L}}{\delta q_i} \delta q_i = \int \frac{\delta \mathcal{L}}{\delta q_i} \left(R_{i\mu} \varepsilon^\mu +  \tilde{R}_{i\mu}^{\ \ \nu}\,\partial_\nu \varepsilon^\mu \right) \nonumber\\
&=& \int \left[\frac{\delta \mathcal{L}}{\delta q_i}\,R_{i\mu} - \partial_\nu\left( \frac{\delta \mathcal{L}}{\delta q_i} \, \tilde{R}_{i\mu}^{\ \ \nu} \right) \right]\varepsilon^\mu = 0,
\label{genNI}
\end{eqnarray}
where the quantity within braces form the Noether identity due to the arbitrary nature of each of the gauge parameters.

The Noether identities corresponding to PGT symmetries are
\begin{eqnarray}
P_k &=& \frac{\delta S}{\delta b^i_{\ \mu}} \epsilon^i_{\ jk} b^j_{\ \mu} + \frac{\delta S}{\delta \omega^i_{\ \mu}} \epsilon^i_{\ jk} \omega^j_{\ \mu} -\partial_\mu\left(\frac{\delta S}{\delta \omega^k_{\ \mu}}\right) = 0 \nonumber\\
R_\rho &=& \frac{\delta S}{\delta b^i_{\ \mu}} \partial_\rho b^i_{\ \mu} + \frac{\delta S}{\delta \omega^i_{\ \mu}} \partial_\rho \omega^i_{\ \mu} - \partial_\mu \left(b^i_{\ \rho}\frac{\delta S}{\delta b^i_{\ \mu}} + \omega^i_{\ \rho} \frac{\delta S}{\delta \omega^i_{\ \mu}} \right) = 0,
\label{PGTNI}
\end{eqnarray}
and that corresponding to canonical hamiltonian symmetries of the Mielke-Baekler action are
\begin{eqnarray}
\!\!A_k \!\!\!\!&=&\!\!\!\! \frac{\delta S}{\delta b^i_{\ \mu}} \epsilon^i_{\ jk} b^j_{\ \mu} + \frac{\delta S}{\delta \omega^i_{\ \mu}} \epsilon^i_{\ jk} \omega^j_{\ \mu} -\partial_\mu\left(\frac{\delta S}{\delta \omega^k_{\ \mu}}\right) \!\!=\! 0 \nonumber\\
\!\!B_k \!\!\!\!&=&\!\!\!\! -\partial_\mu\!\left(\!\frac{\delta S}{\delta b^k_{\ \mu}}\!\right) + \frac{\delta S}{\delta b^i_{\ \mu}} \epsilon^i_{\ jk} \omega^j_{\ \mu} -p \frac{\delta S}{\delta b^i_{\ \mu}} \epsilon^i_{\ jk} b^j_{\ \mu} - q \frac{\delta S}{\delta \omega^i_{\ \mu}} \epsilon^i_{\ jk} b^j_{\ \mu} \!\!=\!\! 0
\label{HamNI}
\end{eqnarray}
A comparison between (\ref{PGTNI}) and (\ref{HamNI}) immediately shows that one identity from each pair is already equivalent: $P_k = A_k$. Comparing the nature of the other identities it is seen that the term $-\omega^k_{\ \rho} A_k -b^k_{\ \rho} B_k$ gives
\begin{eqnarray*}
-R_\rho + \frac{\delta S}{\delta b^i_{\ \mu}} \left(\frac{\alpha_3}{\bigtriangleup}\,\eta^{ij} \epsilon_{\mu\nu\rho}\right) \frac{\delta S}{\delta b^j_{\ \nu}} + \frac{\delta S}{\delta b^i_{\ \mu}} \left(\frac{-a}{\bigtriangleup}\,\eta^{ij}\, \epsilon_{\mu\nu\rho}\right) \frac{\delta S}{\delta \omega^j_{\ \nu}} & \nonumber\\
+  \frac{\delta S}{\delta \omega^i_{\ \mu}}\left(\frac{-a}{\bigtriangleup}\,\eta^{ij}\, \epsilon_{\mu\nu\rho}\right) \frac{\delta S}{\delta b^j_{\ \nu}} + \frac{\delta S}{\delta \omega^i_{\ \mu}}\left(\frac{\alpha_4}{\bigtriangleup}\,\eta^{ij}\, \epsilon_{\mu\nu\rho}\right) \frac{\delta S}{\delta b^j_{\ \nu}} &= 0
\end{eqnarray*}
where $\bigtriangleup = 2\,(\alpha_3\alpha_4 - a^2)$. The terms proportional to square of Euler derivatives are antisymmetric in their co-efficients and as such drop out {\em without} having to use the equations of motion, i.e. without having to set the Euler derivatives to zero. Thus we get back the Poincar\'{e} Noether identities from the canonical hamiltonian Noether identities, their difference being just `trivial' gauge identities \cite{Henneaux}. Thus substituting $R_\rho = -b^k_{\ \rho}B_k - \omega^k_{\ \rho}A_k$ and $P_k = -A_k$ in $\delta S = \int \left( \theta^k P_k + \xi^\rho R_\rho \right) = 0$ gives $\int \left[ (-\theta^k - \xi^\rho\omega^k_{\ \rho})\, A_k\, + \,(-b^k_{\ \rho}\xi^\rho)\, B_k \right] = 0$. Comparing this with $\delta S = \int \left( \varepsilon^k A_k + \tau^k B_k \right) = 0$ gives us the required map between the two sets of gauge parameters.
\begin{equation}
\varepsilon^i = -\xi^\rho b^i_{\ \rho}\quad \& \quad \tau^i=-\theta^i-\xi^\rho\omega^i_{\ \rho}.
\label{map}
\end{equation}
So the Noether identities help us to generate the required map between different sets of gauge parameters and show the equivalence of the two symmetries as their difference is just `trivial!'

\section*{References}

\bibliography{debraj}

\end{document}